\documentclass[11pt]{article}
\usepackage{graphicx}
\usepackage{amsmath}
\input psfig.sty

\newlength{\greater}

\def\q{\textsc{q}}
\def\de{\textsc{de}}
\def\x{\textsc{x}}

\begin{document}

\settowidth{\greater}{ >}

\begin{titlepage}

\title{\bf The Fate of an Accelerating Universe\\
\vspace{0.2cm}}

\author{Je-An Gu\thanks{%
E-mail address: jagu@phys.ntu.edu.tw} \ \ and \ W-Y. P. Hwang\thanks{%
E-mail address: wyhwang@phys.ntu.edu.tw} \\
{\small Department of Physics, National Taiwan University, Taipei
106, Taiwan, R.O.C.}
\medskip
}

\date{\small \today}

\maketitle

\begin{abstract}
The presently accelerating universe may keep accelerating forever,
eventually run into the event horizon problem, and thus be in
conflict with the superstring idea. In the other way around, the
current accelerating phase as well as the fate of the universe may
be swayed by a negative cosmological constant, which dictates a
big crunch. Based on the current observational data, in this paper
we investigate how large the magnitude of a negative cosmological
constant is allowed to be. In addition, for distinguishing the
sign of the cosmological constant via observations, we point out
that a measure of the evolution of the dark energy equation of
state may be a good discriminator. Hopefully future observations
will provide much more detailed information about dark energy and
thereby indicates the sign of the cosmological constant as well as
the fate of the presently accelerating universe.
\end{abstract}




\thispagestyle{empty}

\end{titlepage}

\section{Introduction} \label{Introduction}
The accelerating expansion of the present universe was suggested
by the type Ia supernova (SN Ia) distance measurements
\cite{Perlmutter:1999np,Riess:1998cb} in 1998 and reinforced
recently by updated SN Ia data
\cite{Tonry:2003zg,Knop:2003iy,Barris:2003dq,Riess:2004nr} and
WMAP measurement \cite{WMAP2003} of cosmic microwave background
(CMB). One general conclusion from these measurements and other
CMB observations in recent years
\cite{Sievers:2002tq,Pearson:2002tr,Kuo:2002ua} is that the
universe has the critical density, consisting of $27\%$
pressureless matter and $73\%$ dark energy with a negative
pressure \cite{CMB&SN} (such that $p/\rho < -0.78$
\cite{WMAP2003}).

The most promising candidates to account for dark energy include
(i) a cosmological constant \cite{Lambda models}, (ii) a slowly
evolving scalar field called ``quintessence''
\cite{Caldwell:1998ii}, (iii) the presence of extra dimensions on
top of the well-known $3+1$ space-time
\cite{Deffayet:2001pu,Gu:DE-ED,Burgess:2003&2004,CasimirDE:2004},
and (iv) the modification of gravity
\cite{Carroll:2003wy,Lue:2003ky,Arkani-Hamed:2003uy}. The
existence of a positive cosmological constant is the simplest
candidate for dark energy. The cosmological constant may be
provided by various kinds of matter/energy, such as the vacuum
energy of quantum fields and the potential energy of classical
fields, and could also be originated in the intrinsic geometry.
Nevertheless, Shapiro and Sola \cite{Shapiro:2000zt} show that the
scaling behavior of such cosmological constant, when embedded in
the framework of the standard model of elementary particle
physics, inevitably leads to a cosmological constant of right size
but opposite in sign. As to other dark energy candidates,
quintessence is usually realized by a slowly evolving mode (or
coherent state) of a real scalar field while the possibility of
realizing quintessence by a complex scalar field was also pointed
out in \cite{Gu:2001tr,Boyle:2001du}. Quintessence provides
dynamical negative pressure and could in principle avoid the fine
tuning problem. For example, the``tracker fields'' proposed by
Zlatev \emph{et al.} entail solutions which will join a common
cosmic evolutionary path, regardless of a wide range of initial
conditions \cite{trackQ}. We shall leave the discussions on the
possibility of understanding the accelerating universe via extra
dimensions to other papers of ours \cite{Gu:DE-ED,CasimirDE:2004}.

Another general consensus is that the present universe is thus
already dominated by dark energy and, in light of general
relativity and cosmological principle, there is no way to overrun
the dominance of dark energy over the other familiar form of
matter or energy. It is therefore of fundamental interest to
examine the possibilities resulting an eternally accelerating
universe
\cite{Hellerman:2001yi,Fischler:2001yj,He:2001za,Gonzalez-Diaz:2001ce}.
If the expansion of the universe keeps accelerating forever, the
universe will have to exhibit an event horizon, raising the issue
about the viability of the current description of string theory:
The existence of the event horizon brings challenges to string
theory, such as the construction of suitable observables to
displace the problematic conventional S-matrix in the universe
with an event horizon
\cite{Hellerman:2001yi,Fischler:2001yj,Witten:2001kn}.

However, there is a pitfall in the chain of the arguments leading
to an eternally accelerating universe. That is, there might be
competitive components in our universe and at present only one
such form, say quintessence of some kind, has assumed its control.
For example, it is perfectly justified to assume that a negative
cosmological constant, whose existence may be a natural occurrence
as suggested in
\cite{Shapiro:2000zt,Guberina:2002wt,Kallosh:2003mt,Gu:2001SSB},
could be present but yet to exert its control in the distant
future. The existence of a negative cosmological constant is more
compatible with string theory \cite{Witten:2001kn}. The
possibility of an accelerating universe with a negative
cosmological constant has been considered in \cite{
Guberina:2002wt,Espana-Bonet:2003vk,Shapiro:2004ch} for the
renormalization-group running of the cosmological constant (vacuum
energy) and in \cite{Felder:2002jk,Alam:2003rw,Kallosh:2003bq} for
quintessence.

Thus, after introducing the basics in cosmology in the next
section, we study several simple examples in Sec.\ \ref{Examples}
for demonstrating various possible fates of the presently
accelerating universe which are swayed by a cosmological constant,
in particular, its sign. In Sec.\ \ref{Phenomenology} we
investigate how negative, confronting the constraints from various
astronomical observations, a cosmological constant is allowed to
be, and propose a phenomenological discriminator of the sign of
the cosmological constant. A summary follows in Sec.\
\ref{Summary}.

\section{Basics} \label{Basics}
In standard cosmology, the universe at large scales is described
by the Friedmann-Robertson-Walker (FRW) metric,
\begin{equation}
ds^2 = dt^2 - a^2(t) \left( \frac{dr^2}{1-kr^2} + r^2 d \Omega ^2
\right) , %
\label{FRW metric}
\end{equation}
which respects from the cosmological principle, i.e., homogeneity
and isotropy. In Eq.\ (\ref{FRW metric}), $a(t)$ is the (cosmic)
scale factor, and $k$ can be chosen to be $+1$, $-1$, or $0$,
corresponding to a closed, open, or flat universe, respectively.
Provided that the matter content of the universe is taken to be a
perfect fluid, the Einstein equations yield
\begin{equation}
\left( \frac{\dot{a}}{a} \right) ^2 + \frac{k}{a^2} = \frac{8 \pi
G}{3} \tilde{\rho} = \frac{\Lambda}{3} + \frac{8 \pi G}{3} \rho \; , %
\label{Friedmann eqn}
\end{equation}
\begin{equation}
\frac{\ddot{a}}{a} = - \frac{4 \pi G}{3} (\tilde{\rho} +
3\tilde{p}) = \frac{\Lambda}{3} - \frac{4 \pi G}{3} (\rho + 3p) \; , %
\label{accel eqn}
\end{equation}
and also imply the energy-momentum conservation,
\begin{equation}
d(\tilde{\rho} a^3) = - \tilde{p} d(a^3) \; , %
\label{stress conserv eqn}
\end{equation}
where $\tilde{\rho}$ and $\tilde{p}$ are effective energy density
and pressure of the perfect fluid including the contributions from
a cosmological constant $\Lambda$ and other matter contents.
As indicated in Eq.\ (\ref{accel eqn}), the expansion of the
universe is accelerating if the pressure $\tilde{p}$ is so
negative that
\begin{equation}
\tilde{p} < - \frac{1}{3} \tilde{\rho} < 0 \; , \quad
\mbox{provided } \tilde{\rho} > 0 \; . %
\label{accel condition}
\end{equation}

The universe is usually considered to be composed of various kinds
of perfect fluids with different types of equations of state:
\begin{equation}
p_i = w_i \rho_i \; ,
\end{equation}
where $\rho_i$ and $p_i$ are energy density and pressure of the
$i$-th component, and the equation-of-state factor $w_i$ in
general may depend on energy density $\rho_i$ and time. Assuming
that each (fluid) component evolves independently and each $w_i$
is constant, we obtain, from Eq.\ (\ref{stress conserv eqn}), a
relation between the energy density $\rho_i$ and the scale factor
$a(t)$:
\begin{equation}
\rho_i \propto a^{-3(1+w_i)} \, . %
\label{relation for rho and a(t)}
\end{equation}
This relation implies that the energy density of the component
with a smaller equation-of-state factor $w$ drops more slowly
along with the expansion of the universe. As a result, eventually
the universe will be dominated by the component with the smallest
equation-of-state factor $w$ as long as the universe keeps
expanding at all times.

\section{Examples} \label{Examples}
As one prominent candidate of dark energy, quintessence may play a
crucial role in swaying the fate of the universe. In addition, the
cosmological constant, which entails the smallest
equation-of-state factor ($w=-1$) among ``non-phantom'' ($w \geq
-1$) dark energy sources, may take over to dominate the universe
eventually. Therefore, in this section we will explore how these
two kinds of energy sources will affect the ultimate fate of the
presently accelerating universe.

\subsection*{Quintessence}

We consider a Lagrangian density of a scalar field for the
quintessence as follows:
\begin{equation}
\mathcal{L} = \sqrt{|g|} \left[ \frac{1}{2} g^{\mu \nu } (\partial
_{\mu} \phi ) (\partial _{\nu} \phi ) - V(\phi ) \right] .
\end{equation}
For simplicity, here we consider a universe which at large scales
is homogeneous, isotropic, and spatially flat, and accordingly can
be described by the flat FRW metric ($k=0$) in Eq.\ (\ref{FRW
metric}). Although we focus on this highly symmetric space at
large scales, the small-scale behavior of the field $\phi$ in
general does make contributions to the stress energy of $\phi$,
and accordingly is involved in the calculation of the
energy-momentum tensor. More precisely, when we invoke the
cosmological principle and the Einstein equations of large scales
(corresponding to the FRW metric), the energy-momentum tensor
contributed by a field is the spatially averaged stress energy of
the field (to which the small-scale spatial variations do make
contributions), but not the stress energy of the spatially
averaged field. Thus, regarding the calculation of the large-scale
energy-momentum tensor contributed by a field, in general we
should take the possible small-scale variations of the field into
consideration, instead of ignoring them simply through the
requirement of homogeneity and isotropy of the universe at large
scales. However, in order for the quintessence field $\phi$ to
generate accelerating expansion, the weak spatial dependence of
the field is requisite, as to be explained in the following.

To describe the small-scale behavior and to calculate the stress
energy of a field, in general we should consider a more general
metric instead of the FRW metric. Here, for simplicity, we assume
that the deviation from the background FRW metric is
insignificant, and accordingly we ignore the metric perturbations.
In this case the field equation of the scalar field $\phi$ is
\begin{equation}
\frac{\partial^2 \phi}{\partial t^2} + 3H \frac{\partial
\phi}{\partial t} - \frac{1}{a^2}\nabla ^2 \phi + V'(\phi ) =0 \; ,%
\label{phi field eqn}
\end{equation}
and the energy density and pressure of the quintessence are given
by
\begin{eqnarray}
\rho_{\q} &=& \frac{1}{2} \left( \frac{\partial \phi}{\partial
                 t} \right) ^2 + \frac{1}{2a^2} \left(
                 \nabla \phi \right) ^2 + V(\phi ) \; , \label{phi rho} \\
p_{\q}    &=&  \frac{1}{2} \left( \frac{\partial \phi}{\partial
               t} \right) ^2 - \frac{1}{6a^2} \left(
              \nabla \phi \right) ^2 - V(\phi ) \; . \label{phi p}
\end{eqnarray}
In Eq.\ (\ref{phi field eqn}), the Hubble expansion rate $H$ is
defined as $H \equiv \dot{a}/a$, and $3H (\partial \phi / \partial
t)$ can be regarded as a damping term caused by the expansion of
the universe. Eqs.\ (\ref{phi rho}) and (\ref{phi p}) imply that
the equation-of-state factor $w_{\q}$ (i.e.\ $p_{\q}/\rho_{\q}$)
of the quintessence ranges from $-1$ to $1$.
As mentioned above, in the Einstein equations corresponding to the
FRW metric which describes the universe at large scales, the
energy-momentum tensor involves the spatial average of the above
energy density and pressure. These spatially averaged (over large
enough scales) quantities should be independent of the spatial
coordinates for the requirement of homogeneity and isotropy.
Furthermore, as indicated by Eqs.\ (\ref{phi field
eqn})--(\ref{phi p}), the spatial variation of $\phi$ (that
provides vanishing $\rho_{\q} + 3p_{\q}$) in general can induce
the time variation of $\phi$ that provides positive pressure.
Thus, in order to generate negative enough pressure and the
accelerating expansion, both the weak spatial dependence and the
slow evolution of the quintessence field $\phi$ are requisite.

As summarized in \cite{Hellerman:2001yi,Fischler:2001yj}, there
are various kinds of quintessence models leading to a perpetually
accelerating universe and accordingly exhibiting an event horizon.
For example, for the potential proposed by Ratra and Peebles,
\begin{equation}
V(\phi ) \sim \exp \left( - \sqrt{\frac{3}{2}(\kappa +1)} \phi
\right) , \label{V with const w}
\end{equation}
with $\kappa < -1/3$, we have solutions which will approach to the
equation of state $p=\kappa \rho$ eventually \cite{Ratra:1988rm}.
Consequently this class of potentials with $\kappa < -1/3$ will
generate a perpetually accelerating universe. In addition, for the
potential (originally studied in \cite{Ratra:1988rm})
\begin{equation}
V(\phi ) = \frac{M^{4+n}}{\phi ^n} \, , %
\label{track potential}
\end{equation}
we have ``tracker solutions'' which approach to an equation of
state $p=-\rho$ asymptotically, insuring the eternal acceleration
\cite{trackQ}. The potential in Eq.\ (\ref{track potential}) can
be classified to a wider class of potentials referred to as
``runaway scalar fields'', in which $V$, $V'$, $V''$, $V'/V$, and
$V''/V$ all approach $0$ as $\phi \rightarrow \infty$
\cite{Steinhardt web}. Steinhardt claimed that the runaway scalar
field will guarantee the eventual acceleration of the universe.
Nevertheless, there are other possible quintessence models
providing alternative results
\cite{Fischler:2001yj,Cline:2001nq,Li:2001pk}. For example, for a
potential which drops to a minimum at $\phi_0$ and then becomes
flat for $\phi > \phi_0$, the universe will be dominated by the
kinetic energy after the scalar field passes the minimum, and then
become decelerating with an equation of state $p=\rho$.

There are two more points which should be added. Firstly, the
existence of a nonzero minimum of the quintessence potential
bounded by below can make a contribution to the cosmological
constant. We will categorize the possibly nonzero minimum of the
potential as a part of the cosmological constant, and accordingly
set this minimum to be zero after introducing the cosmological
constant. Secondly, we note that the kinetic energy and potential
energy of the quintessence are transferred to each other along
with the evolution of the quintessence field. In addition, the
kinetic energy is dissipated, accompanying the expansion of the
universe, due to the damping term $3H(\partial \phi / \partial t)$
in Eq.\ (\ref{phi field eqn}). As a result, in general the energy
density of the quintessence will be dissipated as the universe
expands, and the equation-of-state factor $w_{\q}$ can only skim
over or approach $-1$ (rather than halting at that point) and
hence should be always larger than that of a cosmological
constant.

\subsection*{The Cosmological Constant}

Provided that there is no ``phantom'' energy with the equation of
state $w<-1$ \cite{Caldwell:2003vq,Stefancic:2003rc}, the
cosmological constant, with the smallest equation-of-state factor
($w=-1$), will eventually dominate the universe as long as the
universe keeps on expanding at all times. In the following, using
Eqs.\ (\ref{Friedmann eqn}) and (\ref{accel eqn}), we will see how
a positive and a negative cosmological constant will affect the
fate of the universe profoundly in totally different ways.

\subsubsection*{(i) $\Lambda > 0$}
In both the open ($k=-1$) and the flat ($k=0$) case, the universe
will expand forever, and the cosmological constant will take over
eventually such that the expansion will eternally accelerate
thereafter, consequently exhibiting an event horizon. For a closed
($k=1$) universe, the situation is more complicated. It involves a
competition between three terms in the Friedmann equation
(\ref{Friedmann eqn}): the curvature term $k/a^2$, the
cosmological constant term $\Lambda /3$, and the energy density
term $\frac{8 \pi G}{3} \rho$ from the matter contents other than
the cosmological constant. Roughly speaking, if the cosmological
constant has already become predominant before the moment when the
curvature term is comparable to the matter density term, the
universe will then expand forever in an accelerating manner and
exhibit an event horizon. Conversely, if the cosmological constant
is still comparatively small at that moment, the universe will
collapse eventually (a big crunch).

\subsubsection*{(ii) $\Lambda < 0$}
In the case of the negative cosmological constant the universe
will always collapse eventually \cite{Tipler:1976} (see also
\cite{Barrow&Tipler:1988}). We can see, 
from Eq.\ (\ref{Friedmann eqn}), that the universe will start to
collapse at the moment when
\begin{equation}
\frac{k}{a^2} = \frac{\Lambda}{3} + \frac{8 \pi G}{3} \rho \; , %
\label{special collapse condition}
\end{equation}
no matter the universe is open, flat, or closed and how close to
zero the negative cosmological constant might be.

\begin{figure}[h!]
\centerline{\psfig{figure=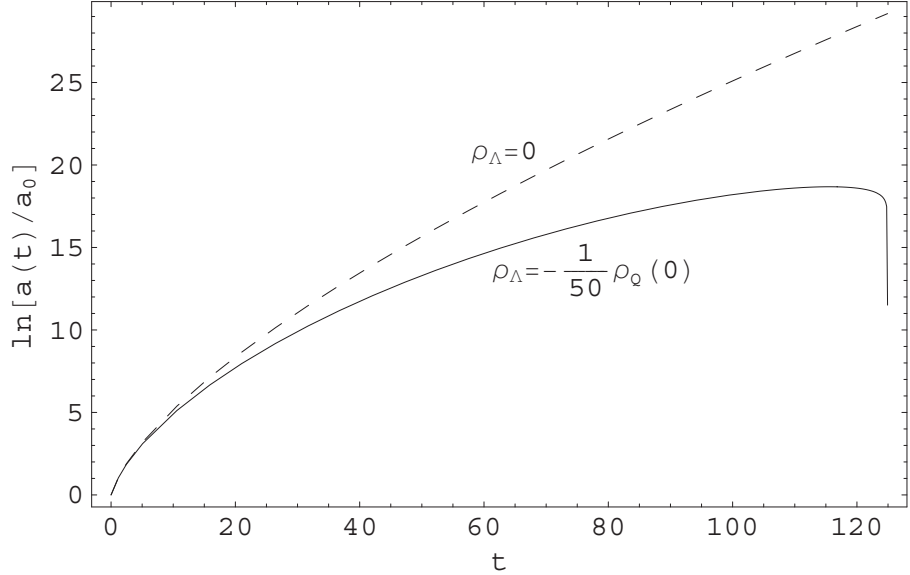,height=3in}} 
\caption{The plot of $ln[a(t)/a_0]$, describing the evolution of
the universe with $\rho_{\scriptscriptstyle
\Lambda}=-\frac{1}{50}\rho_{\q}(0)$ and $\rho_{\scriptscriptstyle
\Lambda}=0$, respectively. The time $t$ is in unit of the Hubble
time $H_0^{-1}$ and $t=0$ denotes the present time.} %
\label{ln-a-plot}
\end{figure}

\begin{figure}[h!]
\centerline{\psfig{figure=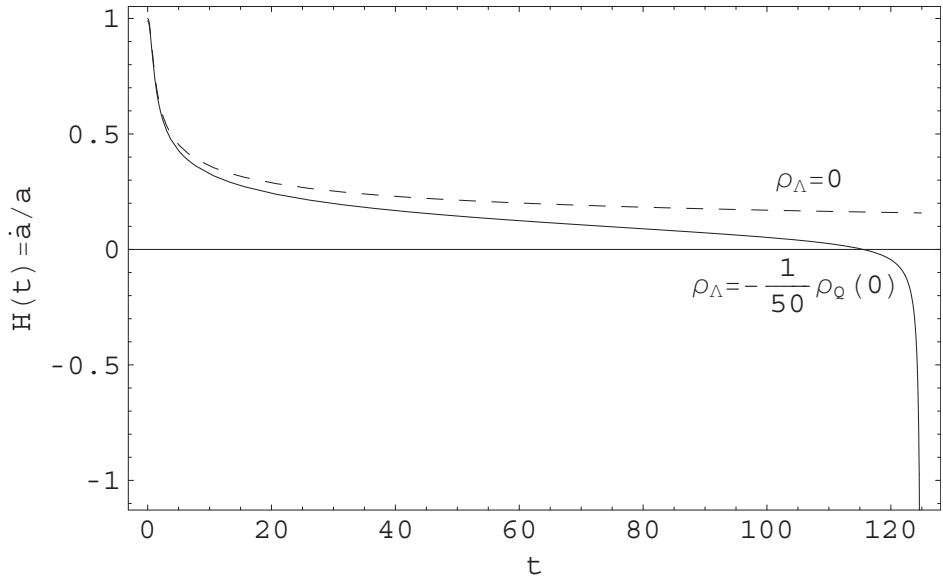,height=3in}} 
\caption{The plot of $H(t)$, describing the evolution of the
universes with $\rho_{\scriptscriptstyle
\Lambda}=-\frac{1}{50}\rho_{\q}(0)$ and $\rho_{\scriptscriptstyle
\Lambda}=0$, respectively. The time $t$ is in unit of the Hubble
time $H_0^{-1}$ and $t=0$ denotes the present time.} %
\label{H-plot}
\end{figure}

For a concrete demonstration, we analyze numerically the evolution
of the universe which is composed of a negative cosmological
constant (with energy density $\rho_{\scriptscriptstyle \Lambda}$)
and a quintessence field $\phi$ with the potential
\begin{equation}
V(\phi ) = \frac{M^6}{\phi ^2} \; . \label{traker demo}
\end{equation}
We use the unit system in which $c = 8 \pi G / 3 = H_0 = 1$ (where
$H_0$ is the present Hubble expansion rate). With this unit
system, the constant $M$ in Eq.\ (\ref{traker demo}) is chosen to
be unity, i.e., $M = (3 H_0 / 8 \pi G)^{1/3} c = 1$. The results
are presented in Figs.\ \ref{ln-a-plot} and \ref{H-plot}. These
two figures sketch the evolution of the scale factor $a(t)$ and
the Hubble expansion rate $H(t)\equiv \dot{a}/a$ of two universes,
where the time $t$ is in unit of the present Hubble time
$H_0^{-1}$ and $t=0$ denotes the present time. The solid line is
for the case of $\rho_{\scriptscriptstyle
\Lambda}=-\frac{1}{50}\rho_{\q}(0)$, where $\rho_{\q}(0)$ is the
present energy density of the quintessence, and the dash line
sketches an accelerating, nearly exponential expansion of the
universe in the case of $\rho_{\scriptscriptstyle \Lambda}=0$ as a
reference. As shown in these two figures, the universe with a
negative cosmological constant will collapse (i.e.\ $\ln
[a(t)/a_0]$ drops and $H(t)$ becomes negative) eventually, in
contrast to the case of $\rho_{\scriptscriptstyle \Lambda}=0$,
even though the energy density from the cosmological constant only
accounts for 2 percent of that from the quintessence initially.

We note that the collapse takes place and proceeds in a violent
manner since the original ``damping'' term $3H(\partial \phi /
\partial t)$ in Eq.\ (\ref{phi field eqn}) turns to an
``amplifying'' term when the universe is collapsing, i.e., $H<0$.
In this collapsing epoch, the kinetic energy $\frac{1}{2}
(\partial \phi / \partial t)^2$ of the quintessence becomes more
and more dominant due to this amplifying effect. After the
universe is dominated by the kinetic energy, it takes a finite
time comparable to or even shorter than the present Hubble time
$H_0^{-1}$ for the universe to collapse to the singularity. This
can be shown by the following formulae for a
quintessential-kinetic-energy-dominated collapsing universe:
\begin{eqnarray}
\left| H \right| %
& \simeq & \sqrt{\frac{8 \pi G}{3} \rho_{\q}}\simeq \sqrt{\frac{4
\pi G}{3} \left( \frac{\partial \phi}{\partial t} \right)^2} \nonumber \\
& > & \sqrt{\frac{8 \pi G}{3} \left| \rho_{\scriptscriptstyle
\Lambda} \right|} \simeq H_0 / \sqrt{50} \; , \\
\left| H \right| %
& \simeq & \frac{1}{3(A-t)} \quad \quad \mbox{($A$ : a constant)} \, . %
\label{collapsing H behavior}
\end{eqnarray}
Equation (\ref{collapsing H behavior}) is straightforwardly
obtained by solving Eqs.\ (\ref{Friedmann eqn}) and (\ref{phi
field eqn}) with the assumption of the dominance of the kinetic
energy.

The above studies reveal a fact that the ultimate fate of the
universe is determined by the cosmological constant (if there is
no ``phantom'' energy). In particular, the future evolution
pattern of our presently accelerating universe under the control
of quintessence can be totally changed by the presence of a
negative cosmological constant, which is part of the dark energy
but may be yet too small to be detected. Consequently, it is
impossible to predict the ultimate fate of our universe without
knowing the detailed nature of the dark energy content therein.

In the following section we shall explore the possibility of the
existence of a negative cosmological constant, confronting the
constraints obtained from various astronomical observations. We
shall see that the negative cosmological constant may not be tiny;
instead, a model with a significant, negative cosmological
constant is consistent with the current observational data.

\section{Phenomenology} \label{Phenomenology}


For simplicity, we consider a flat universe 
dominated by three components: (i) non-relativistic matter,
including (cold) dark matter, with $\Omega_m \cong 0.27$
\cite{Spergel:2003cb}, (ii) a cosmological constant $\Lambda$, and
(iii) a smoothly distributed energy source X that possesses
positive energy density and a constant, negative equation-of-state
factor $w_{\x}$ and is responsible for the present accelerating
expansion. For X to be different from a cosmological constant, we
require $w_{\x} \neq -1$. In this model, dark energy consists of
two components, $\Lambda$ and X; its energy density, pressure, and
the equation-of-state factor are as follows: $\rho_{\de} =
\rho_{\x} + \rho_{\scriptscriptstyle \Lambda}$, $p_{\de} = p_{\x}
+ p_{\scriptscriptstyle \Lambda}$, and
\begin{equation}
w_{\de} = \frac{p_{\x} + p_{\scriptscriptstyle \Lambda}}{\rho_{\x}
          + \rho_{\scriptscriptstyle \Lambda}} %
        = \frac{w_{\x} \rho_{\x} - \rho_{\scriptscriptstyle \Lambda}}
          {\rho_{\x} + \rho_{\scriptscriptstyle \Lambda}} %
\, .
\end{equation}

In this model there are two free parameters: $w_{\x}$ and
$\Omega_{\scriptscriptstyle \Lambda}$ (or
$\rho_{\scriptscriptstyle \Lambda}$). We can obtain information
about these two parameters by invoking the constraints from
observations on various physical quantities. In the following, we
shall invoke the constraints obtained by Wang and Tegmark in
\cite{Wang:2004py}, using SN Ia (in particular, the ``gold'' set
\cite{Riess:2004nr}), CMB
\cite{Pearson:2002tr,Kuo:2002ua,WMAP2003}, and galaxy clustering
\cite{2dFGRS:2002} data. These constraints are respectively on the
dark energy density $\rho_{\de}$ as a function of redshift $z$ and
on the parameters $w_{\de}(0)$ and $w'_{\de}(0)$, where `0'
denotes the present time (i.e., $z=0$), and $w'_{\de}(0) \equiv
dw_{\de}/dz \, (z=0)$.

Introducing $\xi \equiv \rho_{\scriptscriptstyle \Lambda}(0) /
\rho_{\x}(0) = \Omega_{\scriptscriptstyle \Lambda} / \Omega_{\x}$,
we write the formulae for $\rho_{\de}(z)$, $w_{\de}$ and
$w'_{\de}(0)$ as follows:
\begin{eqnarray}
\frac{\rho_{\de}(z)}{\rho_{\de}(0)} &=&
\frac{(1+z)^{3(1+w_{\x})}+\xi}{1+\xi} \, , \\
w_{\de} \hspace{0.5em} &=&
\frac{w_{\x}(1+z)^{3(1+w_{\x})}-\xi}{(1+z)^{3(1+w_{\x})}+\xi} \, ,
\\
1+w_{\de}(0) &=& \frac{1+w_{\x}}{1+\xi} \, , \label{w0} \\
w'_{\de} (0) &=& \frac{3 (1+w_{\x})^2 \xi}{(1+\xi)^2} = 3 \left[ 1
+ w_{\de}(0) \right]^2 \xi \, . \label{w0-prime}
\end{eqnarray}
We require $\xi > -1$ for insuring that the dark energy density is
positive. According to Eq.\ (\ref{w0}), if $\xi > -1$,
$w_{\de}(0)$ and $w_{\x}$ are simultaneously larger or smaller
than $-1$. Accordingly the sign of $1+w_{\de}(0)$ can tell us
whether the dark energy component X is ``phantom'' ($w_{\x} < -1$)
or not. In addition, as shown in Eq.\ (\ref{w0-prime}), the sign
of $w'_{\de}(0)$, i.e., whether the equation-of-state factor of
dark energy is increasing or decreasing, is determined by the sign
of the cosmological constant. As a result, in our model the sign
of $w'_{\de}(0)$ can be a discriminator of the sign of the
cosmological constant phenomenologically. In particular, currently
increasing (i.e., decreasing with $z$) $w_{\de}$ indicates the
existence of a negative cosmological constant.

\begin{figure}[h!]
\begin{minipage}{3in}
\centerline{\psfig{figure=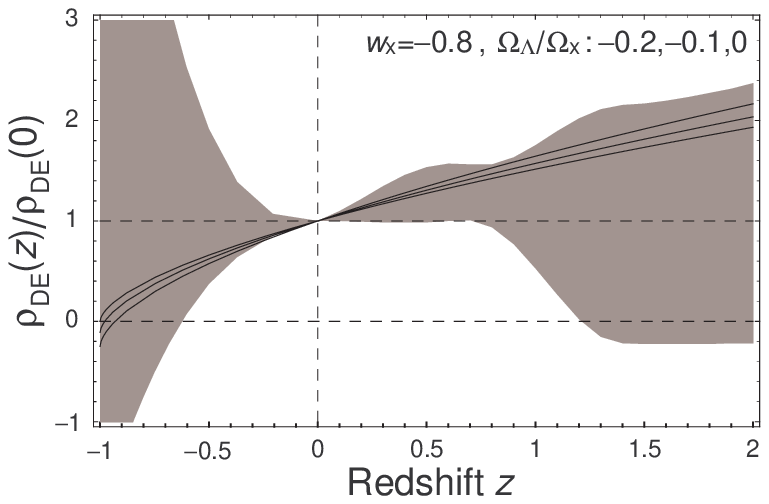,width=3in}}
\end{minipage}
\begin{minipage}{3in}
\centerline{\psfig{figure=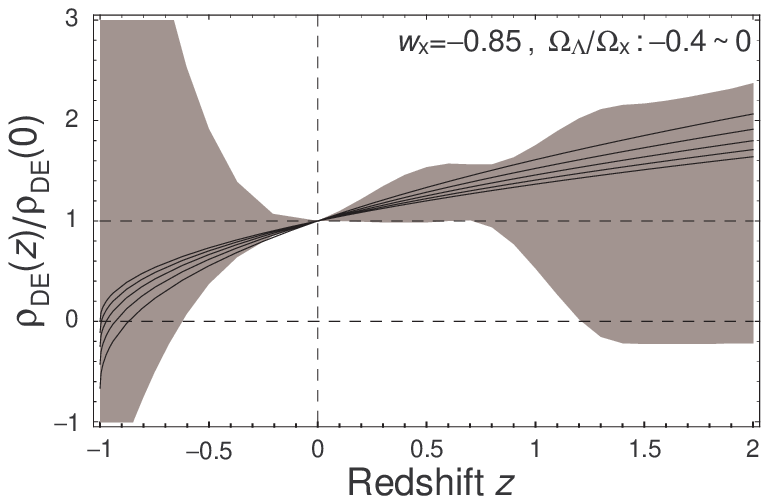,width=3in}}
\end{minipage}
\\
\begin{minipage}{3in}
\centerline{\psfig{figure=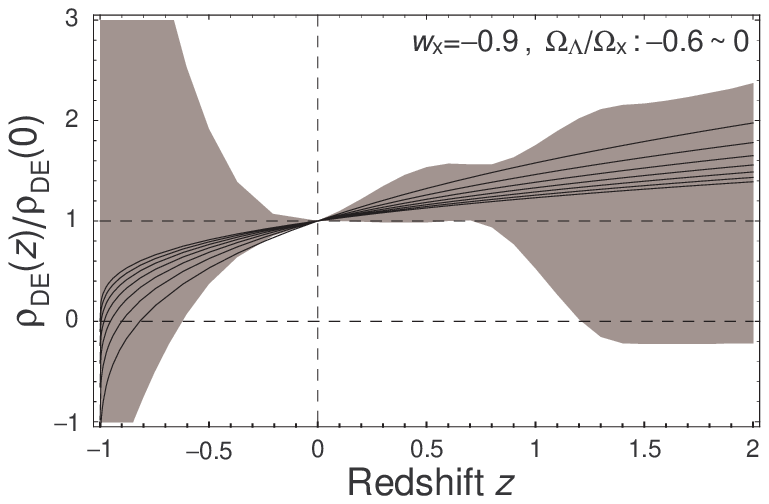,width=3in}}
\end{minipage}
\begin{minipage}{3in}
\centerline{\psfig{figure=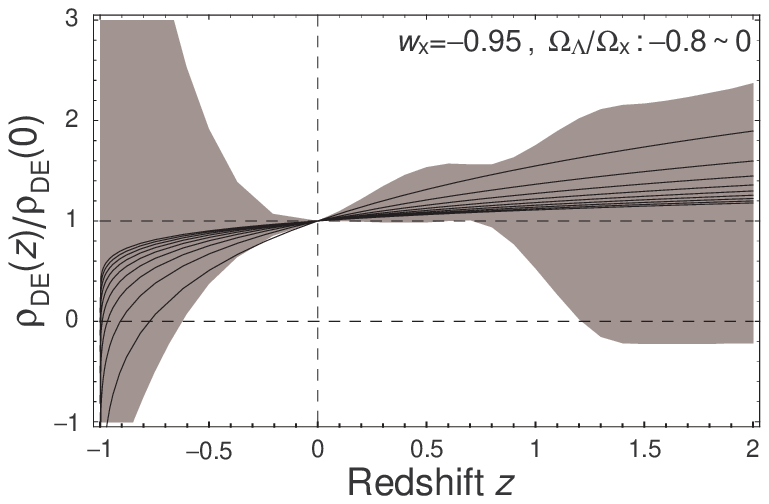,width=3in}}
\end{minipage}
\caption{The $1\sigma$ constraint on $\rho_{\de}(z)$ (presented by
the grey region) and the predictions in various theoretical models
which satisfy this constraint (presented by solid curves).}
\label{constraint on rho} %
\end{figure}

In Fig.\ \ref{constraint on rho}, we demonstrate how well this
model with a negative cosmological constant satisfies the
constraint on the evolution of the dark energy density. The grey
region shows the $1\sigma$ constraint on $\rho_{\de}(z)$, which is
obtained in \cite{Wang:2004py} by using the parametrization of the
3-parameter spline. The solid curves present the behavior of
$\rho_{\de}(z)$ predicted in models with various values of
$w_{\x}$ and $\xi$ which satisfy the $1\sigma$ constraint. These
plots show that the existence of a significant, negative
cosmological constant is allowed. For example, for the case
$w_{\x}=-0.9$, $-\xi$ is allowed to be as large as $0.6$, that is,
at present the magnitude of the energy density of the cosmological
constant is allowed to be larger than one half of that of the dark
energy component X, and accordingly is larger than the dark energy
density.\footnote{Here we consider only $w_{\x}>-1$ and, in
addition, show only curves corresponding to negative $\xi$. Note
that models with larger (positive) $\xi$ are more similar to the
$\Lambda$CDM model that is consistent with all the current data,
and, roughly speaking, are easier to survive the constraints from
observations. Apparently, when $\xi$ goes to infinity, we obtain
the $\Lambda$CDM model with $\rho_{\de}(z)/\rho_{\de}(0)=1$.}

\begin{figure}[h!]
\centerline{\psfig{figure=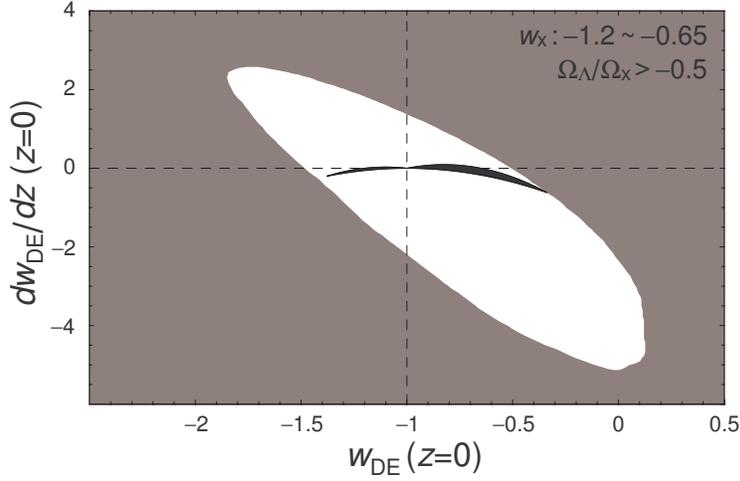,height=2.85in}} 
\caption{The constraint on $w_{\de}(0)$ and $w'_{\de}(0)$
(presented by the white region) and the predictions in various
theoretical models: $-1.2<w_{\x}<-0.65$ and $\xi \equiv
\Omega_{\scriptscriptstyle \Lambda} / \Omega_{\x} > -0.5$
(presented by the black area), which survive this constraint. The
grey region is ruled out by SN Ia data, with the prior
$\Omega_m=0.27 \pm 0.04$, at $95\%$ confidence.}
\label{constraint on w & models 1} %
\end{figure}

\begin{figure}[h!]
\centerline{\psfig{figure=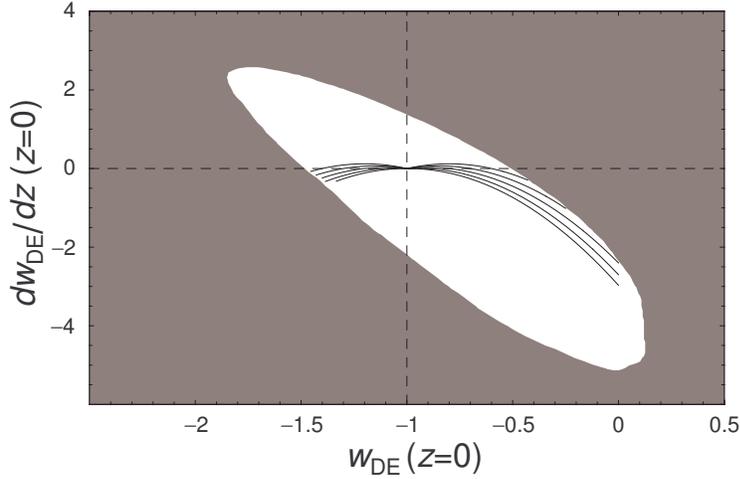,height=2.85in}} 
\caption{The constraint on $w_{\de}(0)$ and $w'_{\de}(0)$
(presented by the white region) and the predictions in various
theoretical models which survive this constraint. The grey region
is ruled out by SN Ia data, with the prior $\Omega_m=0.27 \pm
0.04$, at $95\%$ confidence. The solid curves, from the left to
the right side,
correspond to theoretical models %
$\{ w_{\x} \, , \, \xi > \xi_m \}$ with %
$(w_{\x} \, , \, \xi_m)$ =  %
$(-1.4  \, , -0.12)$ , %
$(-1.3  \, , -0.3 )$ , %
$(-1.2  \, , -0.5 )$ , %
$(-1.1  \, , -0.74)$ , %
$(-1.01 \, , -0.97)$ , %
$(-0.99 \, , -0.99)$ , %
$(-0.9  \, , -0.9 )$ , %
$(-0.8  \, , -0.8 )$ , %
$(-0.7  \, , -0.6 )$ , %
$(-0.6  \, , -0.3 )$.}
\label{constraint on w & models 2} %
\end{figure}

In Figs.\ \ref{constraint on w & models 1} and \ref{constraint on
w & models 2}, we demonstrate how well this model with a negative
cosmological constant satisfies the constraint on $w_{\de}(0)$ and
$w'_{\de}(0)$. The grey region is ruled out by SN Ia data, with
the prior $\Omega_m = 0.27 \pm 0.04$, at 95\% confidence
\cite{Wang:2004py}. In Fig.\ \ref{constraint on w & models 1}, the
black area represents the models with the parameters in the range
$\left\{ -1.2<w_{\x}<-0.65 \, , \, \xi>-0.5 \right\}$, which are
within the allowed (white) region. We note that $\xi=-0.5$
corresponds to $ |\Omega_{\scriptscriptstyle \Lambda}| =
\Omega_{\de} = \Omega_{\x}/2$ and therefore indicates a
significant, negative cosmological constant. For more details, in
Fig.\ \ref{constraint on w & models 2} we illustrate the allowed
range of $\xi$ for various values of $w_{\x}$. From the left to
the right side, the solid curves correspond to theoretical models
$\{ w_{\x} \, , \, \xi > \xi_m \}$ with different values of
$w_{\x}$ and $\xi_m$, as listed in the following:
\begin{center}
\begin{tabular}{c|cccccccccc} 
$w_{\x}$ & $-1.4$ & $-1.3$ & $-1.2$ & $-1.1$ & $-1.01$ & $-0.99$ &
$-0.9$ & $-0.8$ & $-0.7$ & $-0.6$ \\ %
\hline %
$\xi_m$ & $-0.12$ & $-0.3$ & $-0.5$ & $-0.74$ & $-0.97$ & $-0.99$
& $-0.9$ & $-0.8$ & $-0.6$ & $-0.3$ \\ %
\end{tabular}
\end{center}

These plots show that, confronting the current constraints from
observations, the existence of a significant, negative
cosmological constant is allowed. For example, $-\xi$ is allowed
to be as large as 0.9 for the case $w_{\x}=-0.9$, and 0.5 for
$w_{\x}=-1.2$. Even for the case $w_{\x}=-0.6$ that is not
consistent with the $1\sigma$ constraint in Fig.\ \ref{constraint
on rho}, $-\xi$ is still allowed to as large as $0.3$. Thus, the
existence of a negative cosmological constant is fairly consistent
with the current observational data. For the case $w_{\x} > -1$,
the existence of a negative cosmological constant dictates a big
crunch as the fate of the presently accelerating universe, while a
positive cosmological constant dictates eternal acceleration. For
$w_{\x} < -1$, the ``phantom'' dark energy component X dictates a
singular fate in a finite time, so-called ``Big Rip'' (i.e., with
the scale factor $a$ blowing up), before which the expansion of
the universe keeps accelerating
\cite{Caldwell:2003vq,Stefancic:2003rc}.

As mentioned above, in this simple model where the dark energy
component X possesses a constant equation-of-state factor
$w_{\x}$, the signs of $1+w_{\de}(0)$ and $w'_{\de}(0)$ (which
hopefully can be obtained from future observations) respectively
can tell us the signs of $1+w_{\x}$ and $\xi$, i.e., whether the
dark energy component X is ``phantom'' or not and whether the
cosmological constant is positive or negative. This feature and
the relation between the fate of the presently accelerating
universe and the parameters $w_{\x}$ and $\xi$ are summarized in
the following table.

\begin{center}
\begin{tabular}{cc|c} 
$w_{\de}(0)$ and $w_{\x}$ & $w'_{\de}(0)$ and
$\Lambda$ & Fate \\ %
\hline %
$>-1$ & $+$ & accelerating forever \\ %
$>-1$ & $-$ & Big Crunch \\ %
$<-1$ & $+$ & accelerating until Big Rip \\ %
$<-1$ & $-$ & accelerating until Big Rip \\ %
\end{tabular}
\end{center}

For a more general model where $w_{\x}(z)$ is not a constant, the
generalized version of Eqs.\ (\ref{w0}) and (\ref{w0-prime}) is as
follows:
\begin{eqnarray}
1+w_{\de}(0) &=& \frac{1+w_{\x}(0)}{1+\xi} \, , \label{w02} \\
w'_{\de} (0) &=& \frac{w'_{\x}(0)}{1+\xi} + \frac{3
\left[ 1+w_{\x}(0) \right]^2 \xi}{(1+\xi)^2} \nonumber \\
&=& \frac{w'_{\x}(0)}{1+\xi} + 3 \left[ 1 + w_{\de}(0) \right]^2
\xi \, . \label{w0-prime2}
\end{eqnarray}
In this case, the information about the sign of $1+w_{\de}(0)$
obtained from observations can still tell us whether the dark
energy component X is ``phantom'' or not. As to the sign of the
cosmological constant, the situation is more complicated. If
$w_{\x}(z)$ is an arbitrary function, there is no hope to extract
information about $\xi$ or $\rho_{\scriptscriptstyle \Lambda}$
because phenomenologically it is redundant to introduce a
cosmological constant in this case. Nevertheless, for each
category of theoretical models where $w_{\x}(z)$ is further
specified by some theoretical requirements, obtaining information
about the cosmological constant from observations is feasible in
principle. For example, for theoretical models where $w_{\x}(z)$
is determined by $n$ parameters, the information about the
cosmological constant can be extracted from the observational
constraints on $w_{\x}(0)$, $w'_{\x}(0)$, $w''_{\x}(0)$, $\ldots$
, $w^{(n)}_{\x}(0)$ where the superscript `$(n)$' denotes the
$n$-th derivative of $w_{\x}$ with respect to redshift $z$. %
In particular, for models where $w_{\x}$ at the present epoch is
barely dynamical and satisfies $w'_{\x}(0) \ll 3 \left[ w_{\x}(0)
+ 1 \right]^2$, the sign of $w'_{\de}$ can still reflect the sign
of the cosmological constant in the same way as that in the case
of constant $w_{\x}$.

\section{Summary} \label{Summary}

To sum up, we have demonstrated, through simple examples, how the
current accelerating phase as well as the fate of the universe can
be swayed by a negative cosmological constant. A negative
cosmological constant may in fact be a natural occurrence, as
suggested in
\cite{Shapiro:2000zt,Guberina:2002wt,Kallosh:2003mt,Gu:2001SSB}.
We have shown that the existence of a significant, negative
cosmological constant is consistent with the current observational
data, if the present accelerating expansion is generated by
another dark energy component. As a result, the present
accelerating situation does not guarantee the existence of an
event horizon in conflict with the current description of string
theory.

We have pointed out that phenomenologically a measure of the
dynamical behavior of the equation-of-state factor of dark energy
($w_{\de}$) may be a good discriminator of the sign of the
cosmological constant: 
positive (negative) $dw_{\de}/dz \, (0)$ corresponds to a positive
(negative) cosmological constant. In addition, the measure of the
sign of $1 + w_{\de}(0)$ can tell us whether the dark energy
component (if exists) other than a cosmological constant is
``phantom'' or not.
These two discriminators work in the cases where the
equation-of-state factor of the dark energy component X barely
changes with time and the cosmological constant is exactly a
constant. In some other cases they may not work. For example, in
the dark-energy model invoking the renormalization-group running
of the cosmological constant
\cite{Shapiro:2000zt,Shapiro:2000dz,Shapiro:2003ui}, the variable
cosmological constant with $w=-1$ can effectively behave as normal
quintessence ($w>-1$) or phantom dark energy ($w<-1$) in different
situations \cite{Sola:2005et,Sola:2005nh}, and therefore does not
fit the above requirements for the validity of these
discriminators.

So far the observational data are not good enough to tell us the
tendency of the evolution of the dark energy equation of state.
Accordingly both the case of a positive cosmological constant
(alone) and the case of a negative cosmological constant (together
with another dark energy component which provides anti-gravity and
generates the present accelerating expansion) are consistent with
the current data. Hopefully future data (e.g., data from Supernova
Acceleration Probe (SNAP) \cite{SNAP:2005} and Joint Efficient
Dark-energy Investigation (JEDI) \cite{Crotts:2005eu}) will
provide much more detailed information about the evolution of the
dark energy equation of state, and thereby indicate the sign of
the cosmological constant and the fate of the presently
accelerating universe.

\section*{Acknowledgements}
This work is supported in part by the National Science Council,
Taiwan, R.O.C.\ (NSC 93-2811-M-002-056, NSC 93-2112-M-002-047, and
NSC 94-2112-M-002-029) and by the CosPA project of the Ministry of
Education (MOE 89-N-FA01-1-4-3).



\begin{thebibliography}{99}

\bibitem{Perlmutter:1999np}
S.~Perlmutter {\it et al.}  [Supernova Cosmology Project
Collaboration],
Astrophys.\ J.\  {\bf 517}, 565 (1999) [arXiv:astro-ph/9812133].

\bibitem{Riess:1998cb}
A.~G.~Riess {\it et al.}  [Supernova Search Team Collaboration],
Astron.\ J.\  {\bf 116}, 1009 (1998) [arXiv:astro-ph/9805201].


\bibitem{Tonry:2003zg}
  J.~L.~Tonry {\it et al.}  [Supernova Search Team Collaboration],
  Astrophys.\ J.\  {\bf 594}, 1 (2003)
  [arXiv:astro-ph/0305008].

\bibitem{Knop:2003iy}
  R.~A.~Knop {\it et al.}  [The Supernova Cosmology Project Collaboration],
  Astrophys.\ J.\  {\bf 598}, 102 (2003)
  [arXiv:astro-ph/0309368].

\bibitem{Barris:2003dq}
  B.~J.~Barris {\it et al.},
  Astrophys.\ J.\  {\bf 602}, 571 (2004)
  [arXiv:astro-ph/0310843].

\bibitem{Riess:2004nr}
  A.~G.~Riess {\it et al.}  [Supernova Search Team Collaboration],
  Astrophys.\ J.\  {\bf 607}, 665 (2004)
  [arXiv:astro-ph/0402512].


\bibitem{WMAP2003}
  C.~L.~Bennett {\it et al.},
  Astrophys.\ J.\ Suppl.\  {\bf 148}, 1 (2003)
  [arXiv:astro-ph/0302207];
%
  D.~N.~Spergel {\it et al.}  [WMAP Collaboration],
  Astrophys.\ J.\ Suppl.\  {\bf 148}, 175 (2003)
  [arXiv:astro-ph/0302209].



\bibitem{Sievers:2002tq}
  J.~L.~Sievers {\it et al.},
  Astrophys.\ J.\  {\bf 591}, 599 (2003)
  [arXiv:astro-ph/0205387].

\bibitem{Pearson:2002tr}
  T.~J.~Pearson {\it et al.},
  Astrophys.\ J.\  {\bf 591}, 556 (2003)
  [arXiv:astro-ph/0205388].

\bibitem{Kuo:2002ua}
  C.~L.~Kuo {\it et al.}  [ACBAR collaboration],
  Astrophys.\ J.\  {\bf 600}, 32 (2004)
  [arXiv:astro-ph/0212289].


\bibitem{CMB&SN}
A.~Balbi {\it et al.},
Astrophys.\ J.\  {\bf 545}, L1 (2000) [arXiv:astro-ph/0005124];
%
P.~de Bernardis {\it et al.}  [Boomerang Collaboration],
in {\it Proceedings of the CAPP2000 conference}
[arXiv:astro-ph/0011469].



\bibitem{Lambda models}
L.~M.~Krauss and M.~S.~Turner,
Gen.\ Rel.\ Grav.\ {\bf 27}, 1137 (1995) [arXiv:astro-ph/9504003];
%
J.~P.~Ostriker and P.~J.~Steinhardt,
Nature {\bf 377}, 600 (1995);
%
A.~R.~Liddle, D.~H.~Lyth, P.~T.~Viana and M.~White,
Mon.\ Not.\ Roy.\ Astron.\ Soc.\ {\bf 282}, 281 (1996)
[arXiv:astro-ph/9512102].


\bibitem{Caldwell:1998ii}
R.~R.~Caldwell, R.~Dave and P.~J.~Steinhardt,
Phys.\ Rev.\ Lett.\ {\bf 80}, 1582 (1998)
[arXiv:astro-ph/9708069].


\bibitem{Deffayet:2001pu}
  C.~Deffayet, G.~R.~Dvali and G.~Gabadadze,
  Phys.\ Rev.\ D {\bf 65}, 044023 (2002)
  [arXiv:astro-ph/0105068].

\bibitem{Gu:DE-ED}
Je-An~Gu and W-Y.~P.~Hwang,
Phys.\ Rev.\ D {\bf 66}, 024003 (2002) [arXiv:astro-ph/0112565];
%
Je-An~Gu,
in {\it Proceedings of 2002 International Symposium on Cosmology
and Particle Astrophysics (CosPA 2002)}, Taiwan, 2002
[arXiv:astro-ph/0209223];
%
Je-An~Gu, W-Y.~P.~Hwang and J.-W.~Tsai,
Nucl.\ Phys.\ B {\bf 700}, 313 (2004) [arXiv:astro-ph/0403641].

\bibitem{Burgess:2003&2004}
Y.~Aghababaie, C.~P.~Burgess, S.~L.~Parameswaran and F.~Quevedo,
Nucl.\ Phys.\ B {\bf 680}, 389 (2004) [arXiv:hep-th/0304256];
%
C.~P.~Burgess,
Annals Phys.\  {\bf 313}, 283 (2004) [arXiv:hep-th/0402200].

\bibitem{CasimirDE:2004}
  Pisin~Chen and Je-An~Gu,
  arXiv:astro-ph/0409238;
%
  eConf {\bf C041213}, 1110 (2004).


\bibitem{Carroll:2003wy}
  S.~M.~Carroll, V.~Duvvuri, M.~Trodden and M.~S.~Turner,
  Phys.\ Rev.\ D {\bf 70}, 043528 (2004)
  [arXiv:astro-ph/0306438].

\bibitem{Lue:2003ky}
  A.~Lue, R.~Scoccimarro and G.~Starkman,
  Phys.\ Rev.\ D {\bf 69}, 044005 (2004)
  [arXiv:astro-ph/0307034].

\bibitem{Arkani-Hamed:2003uy}
  N.~Arkani-Hamed, H.~C.~Cheng, M.~A.~Luty and S.~Mukohyama,
  JHEP {\bf 0405}, 074 (2004)
  [arXiv:hep-th/0312099].


\bibitem{Shapiro:2000zt}
I.~L.~Shapiro and J.~Sola,
Phys.\ Lett.\ B {\bf 475}, 236 (2000) [arXiv:hep-ph/9910462].


\bibitem{Gu:2001tr}
Je-An~Gu and W-Y.~P.~Hwang,
Phys.\ Lett.\ B {\bf 517}, 1 (2001) [arXiv:astro-ph/0105099].

\bibitem{Boyle:2001du}
L.~A.~Boyle, R.~R.~Caldwell and M.~Kamionkowski,
Phys.\ Lett.\ B {\bf 545}, 17 (2002) [arXiv:astro-ph/0105318].


\bibitem{trackQ}
I.~Zlatev, L.~Wang and P.~J.~Steinhardt,
Phys.\ Rev.\ Lett.\  {\bf 82}, 896 (1999)
[arXiv:astro-ph/9807002];
%
P.~J.~Steinhardt, L.~Wang and I.~Zlatev,
Phys.\ Rev.\ D {\bf 59}, 123504 (1999) [arXiv:astro-ph/9812313].



\bibitem{Hellerman:2001yi}
S.~Hellerman, N.~Kaloper and L.~Susskind,
JHEP {\bf 0106}, 003 (2001) [arXiv:hep-th/0104180].

\bibitem{Fischler:2001yj}
W.~Fischler, A.~Kashani-Poor, R.~McNees and S.~Paban,
JHEP {\bf 0107}, 003 (2001) [arXiv:hep-th/0104181].


\bibitem{He:2001za}
X.~G.~He,
arXiv:astro-ph/0105005.

\bibitem{Gonzalez-Diaz:2001ce}
P.~F.~Gonzalez-Diaz,
Phys.\ Lett.\ B {\bf 522}, 211 (2001) [arXiv:astro-ph/0110335].


\bibitem{Witten:2001kn}
  E.~Witten,
  arXiv:hep-th/0106109.





\bibitem{Guberina:2002wt}
  B.~Guberina, R.~Horvat and H.~Stefancic,
  Phys.\ Rev.\ D {\bf 67}, 083001 (2003)
  [arXiv:hep-ph/0211184].


\bibitem{Kallosh:2003mt}
R.~Kallosh and A.~Linde,
JCAP {\bf 0302}, 002 (2003) [arXiv:astro-ph/0301087].


\bibitem{Gu:2001SSB}
Je-An~Gu and W-Y.~P.~Hwang,
Mod.\ Phys.\ Lett.\ A {\bf 17}, 1979 (2002)
[arXiv:hep-th/0102179];
%
arXiv:hep-th/0105133.




\bibitem{Espana-Bonet:2003vk}
  C.~Espana-Bonet, P.~Ruiz-Lapuente, I.~L.~Shapiro and J.~Sola,
  JCAP {\bf 0402}, 006 (2004)
  [arXiv:hep-ph/0311171].

\bibitem{Shapiro:2004ch}
  I.~L.~Shapiro, J.~Sola and H.~Stefancic,
  JCAP {\bf 0501}, 012 (2005)
  [arXiv:hep-ph/0410095].



\bibitem{Felder:2002jk}
  G.~N.~Felder, A.~V.~Frolov, L.~Kofman and A.~V.~Linde,
  Phys.\ Rev.\ D {\bf 66}, 023507 (2002)
  [arXiv:hep-th/0202017].

\bibitem{Alam:2003rw}
  U.~Alam, V.~Sahni and A.~A.~Starobinsky,
  JCAP {\bf 0304}, 002 (2003)
  [arXiv:astro-ph/0302302].

\bibitem{Kallosh:2003bq}
  R.~Kallosh, J.~Kratochvil, A.~Linde, E.~V.~Linder and M.~Shmakova,
  JCAP {\bf 0310}, 015 (2003)
  [arXiv:astro-ph/0307185].




\bibitem{Ratra:1988rm}
B.~Ratra and P.~J.~Peebles,
Phys.\ Rev.\ D {\bf 37}, 3406 (1988).

\bibitem{Steinhardt web}
P.~J.~Steinhardt, http://feynman.princeton.edu/\~{}steinh/



\bibitem{Cline:2001nq}
J.~M.~Cline,
JHEP {\bf 0108}, 035 (2001) [arXiv:hep-ph/0105251].

\bibitem{Li:2001pk}
M.~Li, W.~Lin, X.~Zhang and R.~H.~Brandenberger,
Phys.\ Rev.\ D {\bf 65}, 023519 (2002) [arXiv:hep-ph/0107160].


%
\bibitem{Caldwell:2003vq}
  R.~R.~Caldwell, M.~Kamionkowski and N.~N.~Weinberg,
  Phys.\ Rev.\ Lett.\  {\bf 91}, 071301 (2003)
  [arXiv:astro-ph/0302506].

\bibitem{Stefancic:2003rc}
  H.~Stefancic,
  Phys.\ Lett.\ B {\bf 586}, 5 (2004)
  [arXiv:astro-ph/0310904].
%


\bibitem{Tipler:1976}
F.~J.~Tipler,
Astrophys.\ J.\  {\bf 209}, 12 (1976).

\bibitem{Barrow&Tipler:1988}
J.~D.~Barrow and F.~J.~Tipler, %
{\it The Anthropic Cosmological Principle} (Oxford University
Press, 1988).


\bibitem{Spergel:2003cb}
  D.~N.~Spergel {\it et al.}  [WMAP Collaboration],
  Astrophys.\ J.\ Suppl.\  {\bf 148}, 175 (2003)
  [arXiv:astro-ph/0302209].

\bibitem{Wang:2004py}
  Y.~Wang and M.~Tegmark,
  Phys.\ Rev.\ Lett.\  {\bf 92}, 241302 (2004)
  [arXiv:astro-ph/0403292].


\bibitem{2dFGRS:2002}
  E.~Hawkins {\it et al.},
  Mon.\ Not.\ Roy.\ Astron.\ Soc.\  {\bf 346}, 78 (2003)
  [arXiv:astro-ph/0212375];
%
  L.~Verde {\it et al.},
  Mon.\ Not.\ Roy.\ Astron.\ Soc.\  {\bf 335}, 432 (2002)
  [arXiv:astro-ph/0112161].


%
\bibitem{Shapiro:2000dz}
  I.~L.~Shapiro and J.~Sola,
  JHEP {\bf 0202}, 006 (2002)
  [arXiv:hep-th/0012227].

\bibitem{Shapiro:2003ui}
  I.~L.~Shapiro, J.~Sola, C.~Espana-Bonet and P.~Ruiz-Lapuente,
  Phys.\ Lett.\ B {\bf 574}, 149 (2003)
  [arXiv:astro-ph/0303306].
%


%
\bibitem{Sola:2005et}
  J.~Sola and H.~Stefancic,
  Phys.\ Lett.\ B {\bf 624}, 147 (2005)
  [arXiv:astro-ph/0505133].

\bibitem{Sola:2005nh}
  J.~Sola and H.~Stefancic,
  arXiv:astro-ph/0507110.
%


\bibitem{SNAP:2005}
  J.~Albert {\it et al.}  [SNAP Collaboration],
  arXiv:astro-ph/0507458;
%
  arXiv:astro-ph/0507460.

\bibitem{Crotts:2005eu}
  A.~Crotts {\it et al.},
  arXiv:astro-ph/0507043.

\end{thebibliography}
\end{document}